\documentclass[a4paper,11pt]{article}
\pdfoutput=1 
\usepackage{jheppub} 
\usepackage[T1]{fontenc} 
\newcommand{\mysmall}[1]{\scriptscriptstyle \rm #1}
\newcommand{ \deltaw}{\delta_{\mysmall{W}}}
\newcommand{ \mw}{M_{\mysmall{W}}}
\newcommand{ \rw}{r_{\mysmall{W}}}
\newcommand{ \gw}{g_{\mysmall{W}}}
\newcommand{ \kw}{k_{\mysmall{W}}}
\newcommand{ \jw}{j_{\mysmall{W}}}
\newcommand{\ymin}{y_0}
\newcommand{\ypmin}{y'_0}
\newcommand{\Lidue}{\textup{Li}_2}

\title{\boldmath Radiative $\mu$ and $\tau$ leptonic decays at NLO}
\author[a]{M. Fael}
\author[b]{L. Mercolli}
\author[c]{M. Passera}
\affiliation[a]{Albert Einstein Center for Fundamental Physics,\\
Institute for Theoretical Physics, University of Bern, CH-3012 Bern, Switzerland}
\affiliation[b]{Princeton University, Department of Astrophysical Sciences, Princeton, NJ, 08544, USA}
\affiliation[c]{Istituto Nazionale Fisica Nucleare, Sezione di Padova, I-35131 Padova, Italy}
\emailAdd{fael@itp.unibe.ch}
\emailAdd{mercolli@astro.princeton.edu}
\emailAdd{passera@pd.infn.it}
\begin{document} 
\abstract{
We present the differential rates and branching ratios of the radiative decays $\tau \to l \bar{\nu} \nu \gamma$, with $l=e$ or $\mu$, and $\mu \to e \bar{\nu} \nu \gamma$ in the Standard Model at next-to-leading order. Radiative corrections are computed taking into account the full depencence on the mass $m_l$ of the final charged leptons, which is necessary for the correct determination of the branching ratios. Only partial agreement is found with previous calculations performed in the $m_l \to 0$ limit. Our results agree with the measurements of the branching ratios $\mathcal{B} (\mu \to e \bar{\nu} \nu \gamma)$ and $\mathcal{B} (\tau \to \mu  \bar{\nu} \nu \gamma)$ for a minimum photon energy of 10~MeV in the $\mu$ and $\tau$ rest frames, respectively. \textsc{Babar}'s recent precise measurement of the branching ratio $\mathcal{B} (\tau \to e \bar{\nu}  \nu \gamma)$, for the same photon energy threshold, differs from our prediction by 3.5 standard deviations.  
}
\maketitle
\flushbottom

\section{Introduction}\label{sec:intro}

Muon and $\tau$ leptonic decays offer one the most powerful tools to study the Lorentz structure of weak interactions. Their theoretical formulation via the Bouchiat-Michel-Kinoshita-Sirlin (BMKS) parameters~\cite{Michel:1949qe,Bouchiat:1957zz,Kinoshita:1957zz,Kinoshita:1957zza} places them in a unique position to investigate possible contributions beyond the $V$--$A$ coupling of the Standard Model (SM). Radiative $\mu$ and $\tau$ leptonic decays, where an inner bremsstrahlung photon is measured, can be predicted with very high precision and provide an independent determination of the BMKS parameters as well as the possibility to extract new combinations like the $\bar{\eta}$ parameter~\cite{Pratt,Eichenberger:1984gi,Fetscher:1993ki}.
A new preliminary measurement of the muon $\bar{\eta}$ was reported recently~\cite{Pocanic:2014mya}, while analyses are ongoing to determine the $\bar{\eta}$ and $\xi \kappa$ parameters of the $\tau$~\cite{Abdesselam:2014uea,Epifanov2014}. 
Precise data on radiative $\tau$ leptonic decays also offer the opportunity to probe the electromagnetic properties of the $\tau$ and may allow to determine its anomalous magnetic moment which, in spite of its precise SM prediction~\cite{Eidelman:2007sb}, has never been measured~\cite{Laursen:1983sm,Fael:2013ij,FaelThesis}.

Recently, the \textsc{Babar} collaboration performed the measurements of the 
$\tau \to l \gamma \nu \bar{\nu} \, (l=e,\mu)$ 
branching fractions for a minimum photon energy $\omega_0=10$~MeV in the $\tau$ rest frame~\cite{Lees:2015gea,OberhofPhDThesis}. The experimental precision of these measurements, around $3\%$, requires the SM prediction of the branching ratios at next-to-leading order (NLO). Indeed these radiative corrections are not protected from mass singularities by the Kinoshita-Lee-Nauenberg (KLN) theorem~\cite{Kinoshita:1958ru,Kinoshita:1962ur,Lee:1964is} and are
therefore expected to be of relative order 
$(\alpha/\pi) \ln(m_l/m_\tau) \ln(\omega_0/m_\tau)$,
corresponding to a large $10\%$ correction for $l=e$, and $3\%$ for $l=\mu$.
Furthermore, special attention must be paid to the role played by the final lepton mass $m_l$ and the limit $m_l \to 0$. In fact, in apparent contradiction to the naive expectation based on the $V$--$A$ weak interaction and helicity conservation in massless QED, final state charged leptons have a finite probability of being right-handed even in the chiral limit $m_l \to 0$. This non-intuitive feature is a consequence of helicity-flip bremsstrahlung in QED and appears as a peculiar mass-singularity cancellation in the collinear region~\cite{Lee:1964is,Falk:1993tf,Sehgal:2003mu,Schulz:2004xd}.

Radiative $\mu$ and $\tau$ leptonic decays also constitute an important source of background for experiments searching for charged lepton flavour violating decays, such as $\mu^+ \to e^+ \gamma$, $\tau \to l \gamma$, and even $\mu^+ \to  e^+ e^- e^+$, because of the internal conversion of photons to electron-positron pairs.
In the next stage of the \textsc{Meg} experiment~\cite{Baldini:2013ke}, as well as in future searches for $\mu^+ \to  e^+ e^- e^+$ at the \textsc{Mu3e} experiment~\cite{Berger:2014vba}, the desired sensitivity will require a refined control and a precise measurement of these backgrounds, which are indistinguishable from the signal except for the missing energy carried away by the neutrinos. An improvement of their theoretical calculation down to a precision of $\mathcal{O}(1\%)$ is desired~\cite{Adam:2013gfn}. Preliminary new measurements of the branching ratio $\mu \to e \gamma \nu \bar{\nu}$ were presented recently by the \textsc{Meg}~\cite{Adam:2013gfn} and \textsc{Pibeta}~\cite{Pocanic:2014mya} collaborations.

We begin our analysis in section~\ref{sec:lo} reviewing the SM prediction for the differential decay rates at leading order (LO). The decay rates at NLO are presented in section~\ref{sec:nlo}. Our NLO predictions for the branching ratios of radiative $\mu$ and $\tau$ leptonic decays are reported in section~\ref{sec:br}, where they are compared with published experimental measurements. Conclusions are drawn in section~\ref{sec:conclusions}.

\section{Differential decay rates at LO}\label{sec:lo}

The LO SM prediction for the differential rates of the radiative decays 
\begin{align}
   \mu^- & \to \, e^- \, \bar{\nu}_e \, \nu_\tau \, \gamma,   
   \label{eqn:muonradiativedecays} \\
   \tau^- & \to \, l^-  \, \bar{\nu}_l \, \nu_\tau \, \gamma,     
   \label{eqn:tauradiativedecays}
\end{align}
with $l=e$ or $\mu$, of a polarized $\mu^-$ or $\tau^-$ in their rest frame is
\begin{multline}
   \frac{d^6 \Gamma_{\mysmall LO}}{dx \, dy \, d\Omega_l\, d\Omega_\gamma}  =
	\frac{\,\alpha G_F^2 M^5} {(4 \pi)^6} 
	\frac{x \beta}{1+ \deltaw(m_{\mu}, m_e)}  \,\, \times \\
	\times \biggl[
	G_{\mysmall LO}(x,y,c)
	+ x \, \beta \, \hat{n} \cdot \hat{p}_l  \, J_{\mysmall LO}(x,y,c) \,\, 
        + y \, \hat{n} \cdot \hat{p}_\gamma \, K_{\mysmall LO}(x,y,c) \biggr],
  \label{eqn:LOdecayrate}
\end{multline}
where 
$G_F=1.166 \, 378 \, 7(6) \times10^{-5}$ GeV$^{-2}$~\cite{PDG2014} 
is the Fermi constant, defined from the muon lifetime, and
$\alpha = 1/137.035\,999\,157\,(33)$
is the fine-structure constant~\cite{Aoyama:2012wj,Aoyama:2014sxa}. Calling $M$ and $m$ the masses of the initial and final charged leptons (neutrinos and antineutrinos are considered massless) we define $r=m/M$ and $\rw=M/\mw$, where $\mw$ is the $W$-boson mass; $p$ and $n=(0,\hat{n})$ are the four-momentum and polarization vector of the initial $\mu$ or $\tau$, with $n^2=-1$ and $n \cdot p = 0$. Also, $x = 2E_l/M$, $y = 2\omega/M$ and $\beta \equiv |\vec{p}_l|/E_l=\sqrt{1-4r^2/x^2}$, where  $p_l = (E_l,\vec{p}_l)$ and $p_\gamma = (\omega,\vec{p}_\gamma)$ are the four-momenta of the final charged lepton and photon, respectively. The final charged lepton and photon are emitted at solid angles $\Omega_l$ and $\Omega_{\gamma}$, with normalized three-momenta $\hat{p}_l$ and $\hat{p}_\gamma$, and  $c \equiv \cos \theta$ is the cosine of the angle between $\hat{p}_l$ and $\hat{p}_\gamma$. The corresponding formula for the radiative decay of a polarized $\tau^+$ or $\mu^+$ is simply obtained inverting the signs in front of the scalar products $\hat{n} \cdot \hat{p}_l$ and $\hat{n} \cdot \hat{p}_\gamma$ in eq.~\eqref{eqn:LOdecayrate}. The term $ \deltaw(m_{\mu}, m_e) =1.04 \times 10^{-6}$ is the tree-level correction to muon decay induced by the $W$-boson propagator~\cite{Ferroglia:2013dga,Fael:2013pja}.

The function $G_{\mysmall LO}$ and, analogously, $J_{\mysmall LO}$ and $K_{\mysmall LO}$, is given by
\begin{equation}
  G_{\mysmall LO} (x,y,c) = \frac{4}{3 y z^2} \Big[ g_0 (x,y,z) \,
  +\, \rw^2 \, \gw (x,y,z) \,+\, {\cal O} \left( \rw^4 \right) \Big],
  \label{eqn:G0}
\end{equation}
where
$
	z = xy \left( 1 - c \beta \right)/2.
$
The functions $g_0$, $j_0$, and $k_0$, computed in refs.~\cite{KSPRL1959,Fronsdal:1959zzb,EcksteinPratt,Fischer:1994pn,Kuno:1999jp}, arise from the pure Fermi $V$--$A$ interaction, whereas $\gw$, $\jw$, and $\kw$, calculated in ref.~\cite{Fael:2013pja}, are the leading contributions induced by the $W$-boson propagator. For the decay  (\ref{eqn:muonradiativedecays}), $\rw^2 \sim 2 \times 10^{-6}$, while for (\ref{eqn:tauradiativedecays}), $\rw^2 \sim 5 \times  10^{-4}$.
Distributions in terms of the helicities of the final lepton and photon were studied in ref.~\cite{Gabrielli:2005ek}. If the initial $\mu^{\pm}$ or $\tau^{\pm}$ are not polarized, eq.~\eqref{eqn:LOdecayrate} simplifies to 
\begin{equation}	
\frac{d^3 \Gamma_{\mysmall LO}}{dx \, dc \, dy}  =
	\frac{\,\alpha G_F^2 M^5} {(4 \pi)^6} \frac{8 \pi^2 \, x \beta}{1+ \deltaw(m_{\mu}, m_e)}  \,\, G_{\mysmall LO}(x,y,c).
\label{eq:radiativedecayrateunpolarized}
\end{equation}

\section{Differential decay rates at NLO}\label{sec:nlo}

We will now consider the SM prediction for the differential rates of the decays (\ref{eqn:muonradiativedecays},\ref{eqn:tauradiativedecays}) at NLO in $\alpha$. These NLO corrections were computed using the effective Fermi Lagrangian, i.e.\ collapsing the SM weak decay vertices, mediated by the $W$-boson, to a four-fermion interaction. In this approximation, tiny terms of $\mathcal{O}(\alpha \, m_{\mu}^2/\mw^2) \sim 10^{-8}$ and $\mathcal{O}(\alpha \, m_{\tau}^2/\mw^2) \sim 10^{-6}$ were neglected, but they are expected to be even smaller than the uncomputed next-to-next-to-leading order (NNLO) corrections of $\mathcal{O}(\alpha^2)$.
Throughout the calculation, the full dependence on the mass ratio $r=m/M$ has been taken into account.

\subsection{QED radiative corrections}

\begin{figure}[htb]
   \centering
   \includegraphics[width=0.433\textwidth]{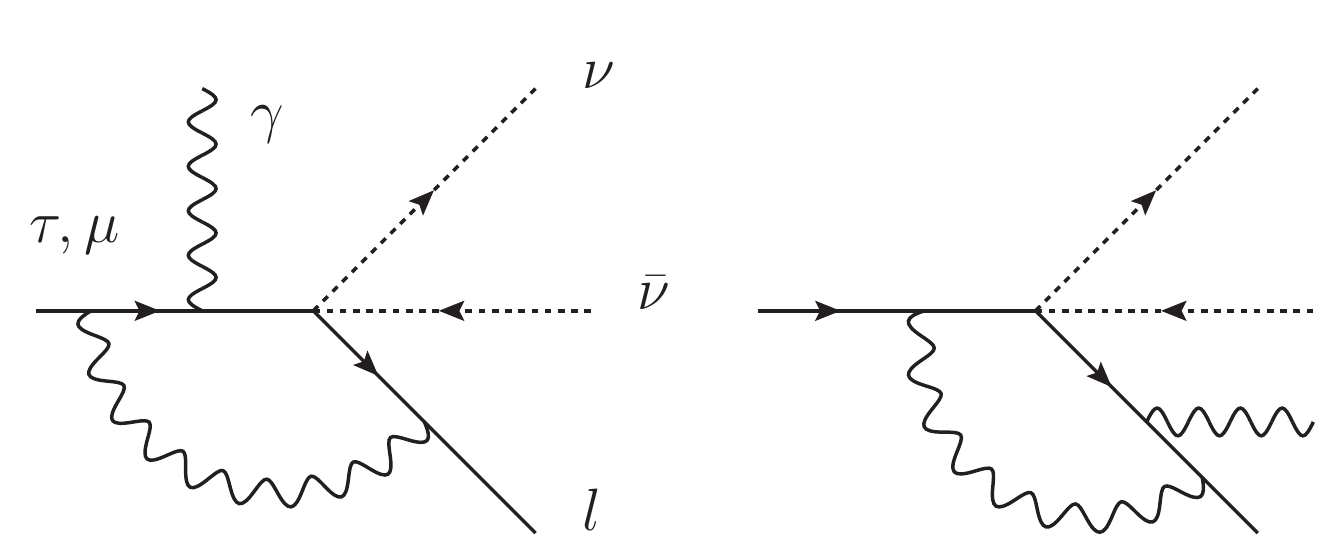}
   \hspace{5mm}
   \includegraphics[width=0.45\textwidth]{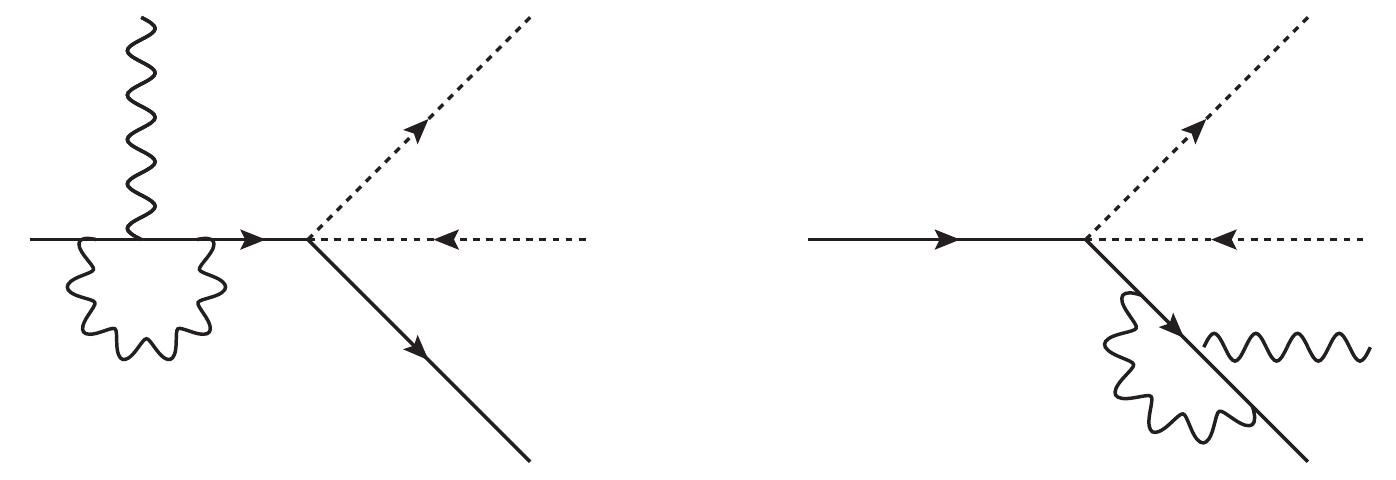}\\
   \vspace{10mm}
   \includegraphics[width=0.428\textwidth]{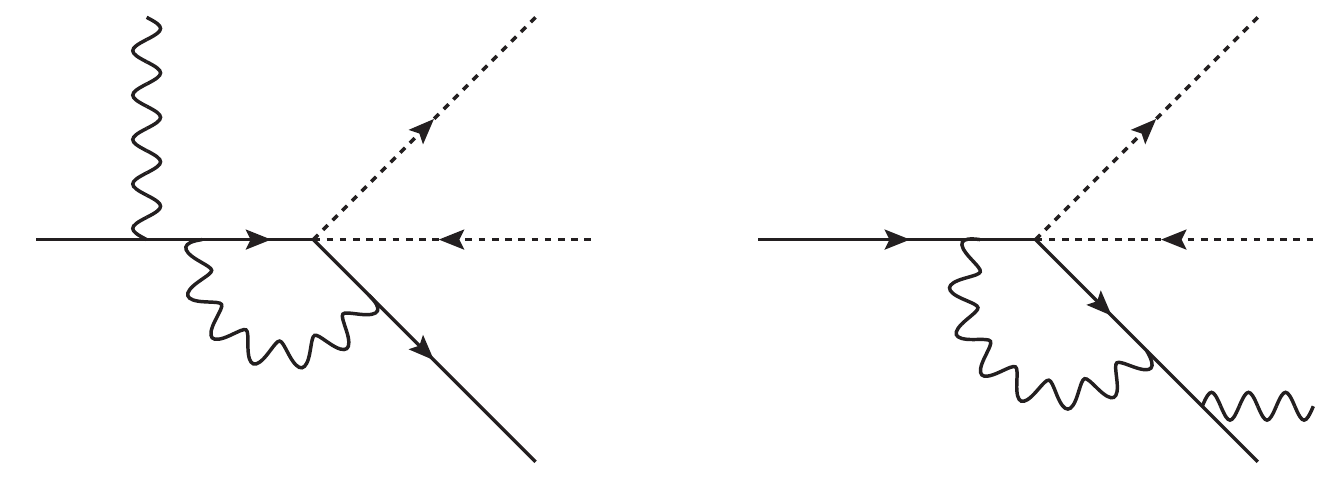}
   \hspace{5mm} 
   \includegraphics[width=0.45\textwidth]{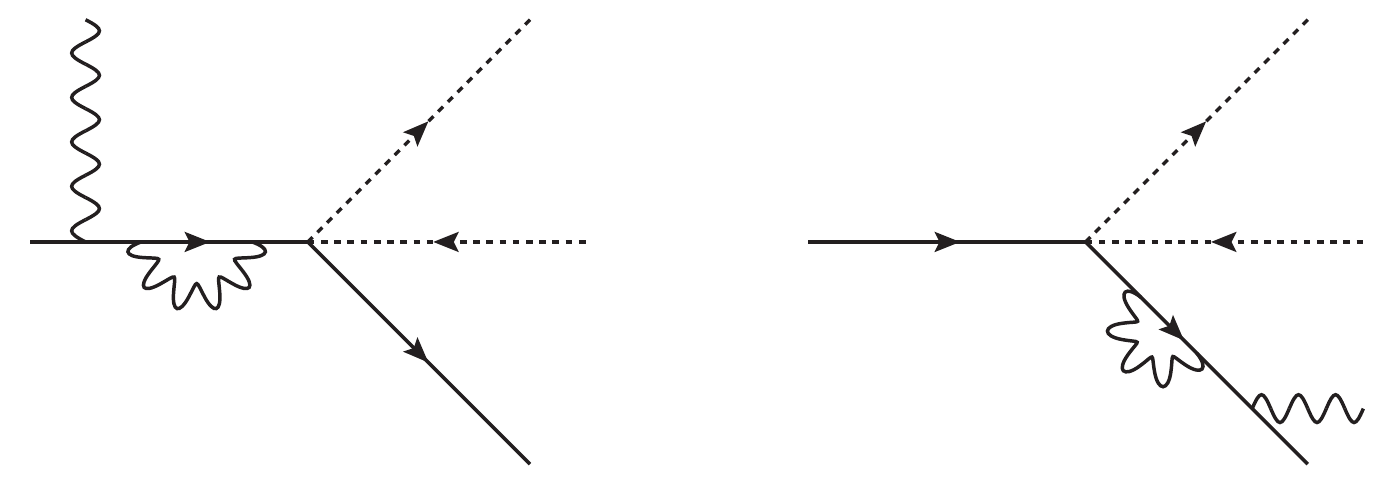}
   \caption{Radiative $\mu$ and $\tau$ leptonic decays: one-loop QED corrections.}
   \label{fig:NLO}
\end{figure}

In the Fermi $V$--$A$ theory, a virtual photon can only be exchanged between charged fermions; see figure~\ref{fig:NLO}. 
We performed the computation of the one-loop QED diagrams via the standard Passarino-Veltman reduction of the tensor integrals~\cite{Passarino:1978jh} using\verb! FORM!~\cite{Kuipers:2012rf} and the \emph{Mathematica} package \verb!FeynCalc!~\cite{Mertig:1990an}. All IR-convergent scalar integrals were computed analytically following~\cite{'tHooft:1978xw} and checked numerically with \verb!LoopTools!~\cite{Hahn:1998yk}; the IR divergent ones were taken from refs.~\cite{Beenakker:1988jr,Ellis:2007qk}. 

UV-finite results were obtained in the on-shell renormalization scheme. Indeed, as shown long ago by Berman and Sirlin~\cite{BermanSirlin1962}, to leading order in $G_F$, but to all orders in $\alpha$, the radiative corrections to muon decay in the Fermi $V$--$A$ theory are finite after mass and charge renormalization. A small photon mass $\lambda$ was introduced to regularize the IR divergences, while the mass ratio $r$ regularized the collinear ones. 
As a check of the calculation of the renormalized one-loop amplitude $\varepsilon^*_\mu(p_\gamma) \mathcal{M}^\mu_{\mysmall virt}$, where $\epsilon^*_\mu(p_\gamma)$ is the polarization vector of the outgoing photon, we explicitly verified that $\mathcal{M}^\mu_{\mysmall virt}$ satisfies the Ward identity
$
      (p_{\gamma })_{\mu} \, \mathcal{M}^\mu_{\mysmall virt}=0.
$

Experimentally, double bremsstrahlung events in which one of the two photons is too soft to be detected are counted as single-photon radiative decays (\ref{eqn:muonradiativedecays},\ref{eqn:tauradiativedecays}). The "soft" differential decay rate for these events should therefore be added to the LO differential decay rate in eq.~\eqref{eqn:LOdecayrate} and to the virtual QED corrections discussed above, thus removing the IR divergence of the latter.

Specifically, let us consider decay events with the emission of two photons with energies $\omega$ and $\omega'$ (and normalized energies $y=2\omega/M$ and $y'=2\omega'/M$), and let us assume that photons can be measured when their normalized energies are above a given threshold $\ypmin$. In the limit $\ypmin \ll 1$, the differential decay rate for the double photon emission obtained integrating $y'$ up to $\ypmin$ is~\cite{Kinoshita:1958ru,Arbuzov:2004wr}
\begin{multline}
   \frac{d^{6} \Gamma_{\gamma \gamma}^{\rm soft}\left(\ypmin\right)}{dx \, dy \, d\Omega_l\, d\Omega_\gamma}  =
	 - \frac{\alpha}{\pi} 	 \left\lbrace  \left( \ln (\ypmin)^2 - \ln \frac{\lambda^2}{M^2} \right)
         \left[1-\frac{1}{2\beta} \ln \left( \frac{1+\beta}{1-\beta} \right) \right] \,\, + \right.  \\
         +\frac{1}{\beta} \textup{Li}_2 \left( \frac{ 2 \beta}{1+\beta} \right)
         - \frac{1}{2\beta} \ln \left( \frac{1+\beta}{1-\beta} \right)
         +\left.\frac{1}{4\beta} \ln^2 \left( \frac{ 1+ \beta}{1-\beta} \right) -1 \right\rbrace
    	\frac{d^6 \Gamma_{\mysmall LO}}{dx \, dy \, d\Omega_l\, d\Omega_\gamma},
\label{eqn:realdecayrateexplicitsoft}
\end{multline}
with $y>\ypmin$. We checked that the IR divergence of this soft bremsstrahlung contribution is canceled by that arising from the virtual QED corrections.

In order to calculate the branching fractions of the radiative decays (\ref{eqn:muonradiativedecays},\ref{eqn:tauradiativedecays}) at NLO (see section~\ref{sec:br}), we also computed the differential rate of the leptonic decay of an unpolarized $\tau$ or $\mu$ with "hard" double-photon emission, which was obtained integrating $y'$ from the threshold $\ypmin$ up to its kinematic upper limit. The calculation of this hard bremsstrahlung differential rate required a detailed analysis of the allowed five-particle phase space which was then integrated numerically with the C{\sc uba} library~\cite{Hahn:2004fe}.

\subsection{NLO results}

The NLO prediction for the differential rate of the radiative $\mu$ and $\tau$ leptonic decays (\ref{eqn:muonradiativedecays},\ref{eqn:tauradiativedecays}), with the possible emission of an additional soft photon with normalized energy $y'<\ypmin$, is
\begin{multline}
   \frac{d^6 \Gamma \left(\ypmin\right) }{dx \, dy \, d\Omega_l\, d\Omega_\gamma}  =
	\frac{\alpha \, G_F^2 M^5} {(4 \pi)^6} 
	\frac{x \beta}{1+ \deltaw(m_{\mu}, m_e)} \,\, \times \\ 
	\times \biggl[
	G
	\, + \, x \beta \, \hat{n} \cdot \hat{p}_l  \, J 
	\, + \, y \, \hat{n} \cdot \hat{p}_\gamma \, K 
        \, + \, x y \beta \, \hat{n} \cdot \left(\hat{p}_l \times \hat{p}_\gamma \right) L
	\biggr],
  \label{eqn:radiativedecayrateNLO}
\end{multline}
with $y>\ypmin$. The function $G (x,y,c; \ypmin)$ and, analogously, $J$ and $K$, is given by
\begin{equation}
  G \, (x,y,c;\ypmin) =
  \frac{4}{3 y z^2} 
  \left[ 
     g_0 (x,y,z) 
     + \rw^2 \, \gw  (x,y,z) 
     + \frac{\alpha}{\pi} \, g_{\mysmall NLO} (x,y,z;\ypmin) 
   \right],
  \label{eqn:GNLO}
\end{equation}
where $g_0 (x,y,z)$ and $\gw  (x,y,z)$ are the LO contributions described in section~\ref{sec:lo}, while $g_{\mysmall NLO} (x,y,z;\ypmin)$ is the sum of the virtual and soft bremsstrahlung contributions~\eqref{eqn:realdecayrateexplicitsoft}. The function $L(x,y,z)$, appearing in front of the term $\hat{n} \cdot \left(\hat{p}_l \times \hat{p}_\gamma \right)$, is only induced by the loop corrections and is therefore of $\mathcal{O}(\alpha/\pi)$. In particular, $L(x,y,z)$ is of the form $\sum_n P_n(x,y,z) \,  {\rm Im} \left[I_n (x,y,z)\right]$, where $P_n$ are polynomials in $x,y,z$ and $I_n (x,y,z)$ are scalar one-loop integrals whose imaginary parts are different from zero. The functions $G$, $J$, $K$ and $L$ are free of UV and IR divergences. Their (lengthy) explicit expressions are provided in attachment to this paper.\footnote{A Fortran code with all the LO and NLO contributions discussed in this paper is available upon request.} If the initial $\mu^{\pm}$ or $\tau^{\pm}$ are not polarized, eq.~\eqref{eqn:radiativedecayrateNLO} simplifies to 
\begin{equation}	
\frac{d^3 \Gamma  \left(\ypmin\right) }{dx \, dc \, dy}  =
	\frac{\,\alpha G_F^2 M^5} {(4 \pi)^6} \frac{8 \pi^2 \, x \beta}{1+ \deltaw(m_{\mu}, m_e)}  \,\, G \, (x,y,c;\ypmin).
\label{eq:radiativedecayrateunpolarizedNLO}
\end{equation}

QED one-loop corrections to radiative muon decays were previously computed in refs.~\cite{Fischer:1994pn,Arbuzov:2004wr} at different levels of completeness, or as part of NNLO corrections to muon decay~\cite{vanRitbergen:1998yd,Anastasiou:2005pn,Caola:2014daa}. The isotropic correction $g_{\mysmall NLO}$ was computed by the authors of ref.~\cite{Fischer:1994pn}, but we couldn't get hold of the files with their expressions. 
%
The authors of ref.~\cite{Arbuzov:2004wr} performed the calculation of radiative $\mu$ decays with the full spin dependence, but in the $r = m_e/m_\mu \to 0$ limit,\footnote{This limit is appropriate if the collinear region is excluded from the phase space integration.} whereas, as we will discuss in section~\ref{sec:br}, terms in $G$ proportional to $r^2$ cannot be neglected if the differential decay rates are integrated over the entire allowed phase space.
We compared our isotropic function $g_{\mysmall NLO}$, taking for this test the limit $r \to 0$, with the result $g_{\mysmall NLO} |_{r\to 0}$ of ref.~\cite{Arbuzov:2004wr}, finding perfect numerical agreement. On the contrary, our results for the anisotropic contributions $j_{\mysmall NLO}$ and $k_{\mysmall NLO}$ differ, even in the $r \to 0$ limit, from those of ref.~\cite{Arbuzov:2004wr}.
Moreover, the contribution of the function $L(x,y,z)$ has been previously overlooked.

\section{Branching Ratios}\label{sec:br}

The kinematic limits of integration for the variables $x$, $c$, and $y$ are
\begin{equation} \label{eqn:fullrange}
	2r \leq  x  \leq 1+r^2, \quad
	-1 \leq  c  \leq 1, \quad
	0  < y \leq y_\textup{max}(x,c), 
\end{equation}
where the maximum normalized photon energy as a function of $x$ and $c$ is
\begin{equation}
	y_{\max}(x, c) = \frac{2 \left(1+r^2-x \right)}{2-x + c  \, x \beta}.
\end{equation}
However, every experimental setup has a minimum photon energy $\omega_0 = \ymin \, (M/2)$ below which photons are not detected. As the constraint $\ymin < y_{\max}(x, c)$, necessary to measure radiative decays, leads to the bound $c < c_{\max}(x)$, with
\begin{equation}
	c_{\max}(x)= \frac{2 \left(1+r^2-x \right) - \big(2-x\big) \ymin }{x \, \beta \, \ymin },
\label{eqn:cmax}
\end{equation}
the kinematic ranges of $x$, $c$, and $y > \ymin$ are reduced to
\begin{equation} \label{eqn:restrictedrange}
	2r \leq  x  \leq 1+r^2, \quad
	-1 \leq  c  \leq \min \{1,c_{\max}(x)\}, \quad
	 \ymin \leq  y  \leq y_{\max}(x,c).
\end{equation}
Integrating the differential rates in eq.~\eqref{eq:radiativedecayrateunpolarizedNLO} over the kinematic ranges \eqref{eqn:restrictedrange} and multiplying them by the $\mu$ or $\tau$ lifetimes
$\tau_{\mu}= 2.1969811(22) \times 10^{-6}$~s
and
$\tau_{\tau}=2.903(5) \times 10^{-13}$~s~\cite{PDG2014},
we obtain the NLO predictions for the branching ratios of the radiative decays (\ref{eqn:muonradiativedecays},\ref{eqn:tauradiativedecays}).

\subsection{Branching ratios at LO}

The analytic integration over the kinematic ranges \eqref{eqn:restrictedrange} of the LO differential rate in eq.~\eqref{eq:radiativedecayrateunpolarized} with a minimum energy $\ymin=2 \omega_0/M$ gives~\cite{EcksteinPratt,KSPRL1959}
\begin{align}
   \Gamma_{\mysmall LO}  \left( \ymin \right) & = 
   \, \frac{G_F^2 M^5}{192 \pi^3}
   \frac{\alpha}{3\pi} \, H\left(\ymin\right),
   \label{eqn:totalLOr} \\
	H\left(\ymin\right) & =   
   \, 3 \, \Lidue (\ymin)  - \frac{\pi^2}{2}  
   + \left( \ln r +\frac{17}{12} \right) 
   \left( 6 \ln \ymin +6 \bar{y}_0 + \bar{y}_0^4 \right)   \, +
		 \notag \\
    &   \, +\, \frac{1}{48} \left(125+45 \ymin -33 \ymin^2 +7 \ymin^3 \right) \bar{y}_0 
    \, - \, \frac{1}{2} \left( 6+\bar{y}_0^3 \right) \bar{y} _0  \ln \bar{y}_0,
\label{eqn:Iy0}
\end{align}
where $\bar{y}_0 = 1-\ymin$ and the dilogarithm is defined by
$\Lidue(z) = -\int_0^z \!dt \,\frac{\ln(1-t)}{t}.$
Terms depending on the mass ratio $r$ have been neglected in the expression for $H(\ymin)$, with the obvious exception of the logarithmic contribution which diverges in the limit $r\to0$. However, as already mentioned in section~\ref{sec:nlo}, terms in the integrand $G_{\mysmall LO} (x,y,c)$ proportional to $r^2$ were not neglected when performing the integral to obtain~\eqref{eqn:Iy0}, as they lead to terms of $\mathcal{O}(1)$ in the integrated result $H(\ymin)$. In fact, the functions multiplying these $r^2$ terms in the integrand generate a singular behavior in the $r \to 0$ limit after the integration over $c \equiv \cos \theta$: terms proportional to $r^2/z^2$ in $G_{\mysmall LO} (x,y,c)$ lead to a nonvanishing contribution to the integrated decay rate since $\int \!  dc \, (1/z^2) \propto 1/z$ is evaluated at the integration limit $c \to 1$ where $z \to xy \left( 1 - \beta \right) \!/2  \approx r^2 (y/x)$ for $x \gg 2r$. This feature, first noted in~\cite{Lee:1964is}, is due to the appearance of right-handed electrons and muons in the final states of (\ref{eqn:muonradiativedecays},\ref{eqn:tauradiativedecays}) even in the limit $r \to 0$, and is a consequence of helicity-flip bremsstrahlung in QED~\cite{Lee:1964is,Falk:1993tf,Sehgal:2003mu,Schulz:2004xd}.
We also note that the presence of the mass singularity $\ln r$ in the integrated decay rate $\Gamma_{\mysmall LO}  \left( \ymin \right)$ does not contradict the KLN theorem, which applies only to total decay rates~\cite{Kinoshita:1958ru, Kinoshita:1962ur, Lee:1964is}.
The tiny corrections induced by the $W$-boson propagator were neglected in~eq.~\eqref{eqn:totalLOr}.

If we multiply the analytic result for $\Gamma_{\mysmall LO} \left( \ymin \right)$ in eq.~\eqref{eqn:totalLOr} by the lifetimes $\tau_{\mu,\tau}$ with a threshold $\omega_0 = 10$~MeV we obtain the following LO predictions for the branching ratios:
$1.83 \times 10^{-2}~(\tau \to e \bar{\nu} \nu \gamma)$, 
$3.58 \times 10^{-3}~(\tau \to \mu \bar{\nu} \nu \gamma)$, and
$1.31 \times10^{-2}~(\mu \to e \nu \bar{\nu}\gamma)$.
These values are in perfect agreement with the results $\mathcal{B}_{\scriptscriptstyle \rm LO}$ reported in table~\ref{tab:BR}, obtained integrating numerically the LO differential rates in eq.~\eqref{eq:radiativedecayrateunpolarized}, with the exception of the $\tau \to \mu \bar{\nu} \nu \gamma$ value; this difference is due to the terms neglected in the analytic result~\eqref{eqn:totalLOr}. We note that, had we performed the numerical integrations of the LO differential rates neglecting terms proportional to $r^2$ in the integrand $G_{\mysmall LO} (x,y,c)$, we would have obtained the incorrect results
$1.96 \times 10^{-2}~(\tau \to e \bar{\nu} \nu \gamma)$,
$4.92 \times 10^{-3}~(\tau \to \mu \bar{\nu} \nu \gamma)$, and 
$1.44 \times10^{-2}~(\mu \to e \nu \bar{\nu}\gamma$).

\subsection{Branching ratios at NLO}

The branching ratios of the radiative decays (\ref{eqn:muonradiativedecays},\ref{eqn:tauradiativedecays}) can be distinguished in two types:
the "exclusive" ones, $\mathcal{B}^{\rm Exc} \left( y_0 \right)$, are measured counting the decays with the emission of one, and only one, photon of energy larger than the detection threshold $y_0$. Such measurements clearly include double bremsstrahlung events in which one of the two photons is too soft to be detected, but decay events with the emission of two hard photons, both with energies larger than $y_0$, must be discarded.
The "inclusive" branching ratios $\mathcal{B}^{\rm Inc} \left( y_0 \right)$ are defined measuring radiative decays with the emission of at least one photon with energy higher than $\ymin$. In these inclusive measurements, if two photons with energies $y$ and $y'$ larger than $y_0$ are detected, the event must be included.

At LO, the theoretical predictions for these exclusive and inclusive branching ratios clearly coincide -- double bremsstrahlung events are simply not considered. 
At NLO, the theoretical prediction for the exclusive branching ratios can be calculated integrating the differential rate $d^3 \Gamma  \left(\ypmin\right)$ in eq.~\eqref{eq:radiativedecayrateunpolarizedNLO}, multiplied by the lifetime, setting $\ypmin = \ymin$. As $d^3 \Gamma  \left(\ypmin\right)$ was obtained adding the analytic formulae for the LO contribution, the virtual corrections, and the soft term $d^{3} \Gamma_{\gamma \gamma}^{\rm soft}\left(\ypmin\right)$ (see eq.~\eqref{eqn:realdecayrateexplicitsoft}), thus cancelling the IR divergences, this result will be called 
\begin{equation}
\mathcal{B}_{\mysmall LVS} (\ymin,\ypmin) = \mathcal{B}_{\mysmall LO} (\ymin) + \mathcal{B}_{\mysmall VS} (\ymin,\ypmin),
\label{eqn:BLVS}
\end{equation}
with $\ypmin = \ymin$, where the LO term has been separated for convenience.\footnote{Note that, in general, the two thresholds $\ymin$ and $\ypmin$ can be different, with $\ypmin \leq \ymin$.}
However, as $d^{3} \Gamma_{\gamma \gamma}^{\rm soft}\left(\ypmin\right)$ was calculated analytically in the limit $\ypmin \ll 1$, $\mathcal{B}_{\mysmall LVS} (\ymin,\ymin)$ is inadequate to predict the exclusive branching ratios if $\ymin$ is large. This is the case, for example, for $\mu \to e \nu \bar{\nu}\gamma$ decays with $\omega_0 = 10$~MeV ($\ymin \sim 0.2$), where the NLO correction $\mathcal{B}_{\mysmall VS} (\ymin,\ymin) = -1.1 \times 10^{-4}$ obtained in this manner turns out to be off by a factor of two w.r.t.\ the correct value $-2.2 \times 10^{-4}$ (see $\mathcal{B}_{\mysmall NLO}^{\rm Exc} \left(\ymin \right)$ later on). We will therefore proceed differently, deriving the exclusive branching ratios from the inclusive ones.

The inclusive branching ratios can be calculated adding to $\mathcal{B}_{\mysmall LVS} (\ymin,\ypmin)$ the branching ratios for the hard double bremsstrahlung events. More precisely, let us call $\mathcal{B}_{\gamma \gamma}^{\rm hard}(\ymin,\ypmin)$ the branching ratio of hard double bremsstrahlung decays where both photons have energies larger than $\ypmin$, and at least one of them has energy above $\ymin$. The NLO prediction for the inclusive branching ratio is given by
\begin{equation}
	\mathcal{B}^{\rm Inc} (\ymin) = \lim_{\ypmin \to 0} \left[  
	\mathcal{B}_{\mysmall LVS} (\ymin,\ypmin)  + \mathcal{B}_{\gamma \gamma}^{\rm hard}(\ymin,\ypmin) \right].
\label{eqn:BINCL}
\end{equation}
The sum in square brackets of eq.~\eqref{eqn:BINCL} should not depend on $\ypmin$, because the sum of soft and hard double-photon emission describes events with one photon of energy larger than $\ymin$ plus another photon of any energy allowed by the kinematics. Therefore, the $\ypmin$-dependent terms of the soft radiation contribution should be canceled by those of the hard one. This, however, is not exactly the case because, as mentioned earlier, the soft term was calculated in the $\ypmin \ll 1$ limit and the cancellation of the $\ypmin$-dependent terms is therefore not exact. 
For this reason we introduced the limit $\ypmin \to 0$ in eq.~\eqref{eqn:BINCL}.

Once the inclusive branching ratios are calculated, the exclusive ones can be computed subtracting 
\begin{equation}
	\mathcal{B}^{\rm Exc} \left(\ymin \right) =   \mathcal{B}^{\rm Inc} \left(\ymin \right) - \mathcal{B}_{\gamma \gamma}^{\rm hard}\left(\ymin,\ymin \right).
\label{eqn:BEXCL}
\end{equation}
The second term on the r.h.s.\ of eq.~\eqref{eqn:BEXCL} subtracts in fact those events in which both photons have energies above the threshold $\ymin$. Such events are discarded in exclusive measurements of the branching fractions. We point out that $\mathcal{B}^{\rm Exc} (\ymin)$ differs, in general, from the branching ratio $\mathcal{B}_{\mysmall LVS} (\ymin,\ymin)$ discussed above (we already mentioned the discrepancy for radiative $\mu$ decays with $\omega_0=10$~MeV). For convenience, $\mathcal{B}^{\rm Exc} (\ymin)$ and $\mathcal{B}^{\rm Inc} (\ymin)$ are separated into LO contributions and NLO corrections:
\begin{align}
   \mathcal{B}^{\rm Exc} \left(\ymin \right) & \, = \,  \mathcal{B}_{\mysmall LO} \left(\ymin \right)  \, + \,
   							\mathcal{B}_{\mysmall NLO}^{\rm Exc} \left(\ymin \right),
   \label{eqn:BEXCLLONLO} \\
   \mathcal{B}^{\rm Inc} \left(\ymin \right) & \, = \,   \mathcal{B}_{\mysmall LO} \left(\ymin \right) \, + \,
   							\mathcal{B}_{\mysmall NLO}^{\rm Inc} \left(\ymin \right).
   \label{eqn:BINCLLONLO}
\end{align}
%

\subsection{Numerical results}

Exclusive and inclusive branching ratios for the radiative decays (\ref{eqn:muonradiativedecays},\ref{eqn:tauradiativedecays}) are presented in table~\ref{tab:BR} for a threshold $\omega_0 = \ymin \, (M/2) = 10$~MeV.
All branching ratios were computed keeping into account the full dependence on the mass ratio $r$. In fact, also at NLO, for the same reasons discussed in the LO case, terms in the integrand $G(x,y,c)$ proportional to $r^2$ cannot be neglected when performing the integrals to calculate the branching ratios.
Numerical integrations were performed with the C{\sc uba} library~\cite{Hahn:2004fe} and tested with different numerical integration methods. 
The hard double bremsstrahlung terms $\mathcal{B}_{\gamma \gamma}^{\rm hard}(\ymin,\ypmin)$, necessary both for $\mathcal{B}^{\rm Exc} (\ymin)$ and for $\mathcal{B}^{\rm Inc} (\ymin)$, required a detailed study of the five-particle phase space which was then integrated numerically. The values obtained for $\mathcal{B}_{\gamma \gamma}^{\rm hard}(\ymin,\ypmin)$ were checked with MadGraph5~\cite{Alwall:2014hca}.

Uncertainties were estimated for uncomputed NNLO corrections, numerical errors, and the experimental errors of the lifetimes.
The former were estimated to be  
$\delta \mathcal{B}^{\rm Exc/Inc}_{\mysmall NLO} \! \sim (\alpha/\pi) \ln r \ln (\omega_0/M) \, \mathcal{B}_{\mysmall NLO}^{\rm Exc/Inc} \!.$
For $\omega_0 = 10$~MeV they are about
10\%, 3\%, and 3\%
for $\tau \to e \bar{\nu} \nu \gamma$, $\tau \to \mu \bar{\nu} \nu \gamma$, and $\mu \to e \nu \bar{\nu}\gamma$,
respectively. They appear with the subscript "$N$" in table~\ref{tab:BR}. The branching ratios due to hard triple-photon emission, estimated with MadGraph5, are much smaller than these uncertainties.
Numerical errors, labeled in table~\ref{tab:BR} by the subscript "$n$", are smaller than those induced by missing radiative corrections. These two kinds of uncertainties were combined to provide the theoretical error of the final $\mathcal{B}^{\rm Exc}$ and $\mathcal{B}^{\rm Inc}$ predictions, labeled  in table~\ref{tab:BR} by the subscript "${\rm th}$". The uncertainty due to the experimental error of the lifetimes is labeled by the subscript "$\tau$"; it is negligible in radiative $\mu$ decays.

The recent measurements by the \textsc{Babar} collaboration of the branching ratios of the radiative decays  $\tau \to l \bar{\nu} \nu \gamma$, with $l=e$ and $\mu$, for a minimum photon energy $\omega_0=10$~MeV in the $\tau$ rest frame, are~\cite{Lees:2015gea,OberhofPhDThesis}:
\begin{align}
   \mathcal{B}_{\scriptscriptstyle \rm EXP} \left(\tau \to e  \bar{\nu}  \nu \gamma  \right) 
   & \, = 1.847  \, (15)_{\rm st} (52)_{\rm sy} \times 10^{-2},
   \label{eqn:TAUEBabar} \\
  \mathcal{B}_{\scriptscriptstyle \rm EXP} \left(\tau \to \mu  \bar{\nu}  \nu \gamma  \right) 
  & \, = 3.69  \, (3)_{\rm st} (10)_{\rm sy} \times 10^{-3},
   \label{eqn:TAUMUBabar}
\end{align}
where the first error is statistical and the second is systematic. These results are substantially more precise than the previous measurements
$1.75(6)_{\rm st} (17)_{\rm sy} \times 10^{-2}~\left(\tau \to e  \bar{\nu}  \nu \gamma  \right)$ and
$3.61(16)_{\rm st} (35)_{\rm sy} \times 10^{-3}~\left(\tau \to \mu  \bar{\nu}  \nu \gamma  \right)$
of the \textsc{Cleo} collaboration~\cite{Bergfeld:1999yh}. 
The signature for $\tau \to l \bar{\nu} \nu \gamma$ decays is a charged particle and a photon. In particular, in the \textsc{Babar} measurements each $e^+ e^- \to \tau^+ \tau^-$ event was divided into hemispheres (signal and tag) in the CM frame. Depending on the signal mode, either a muon or an electron with a single photon candidate were required on the signal side~\cite{Lees:2015gea,OberhofPhDThesis}. The experimental values in eqs.~(\ref{eqn:TAUEBabar},\ref{eqn:TAUMUBabar}) should therefore be compared with our predictions for the exclusive branching ratios
$ 1.645  \, (19)_{\rm th} (3)_{\rm \tau} \times 10^{-2}$ and 
$ 3.572  \, (3)_{\rm th} (6)_{\rm \tau} \times 10^{-3}$,
respectively (see table~\ref{tab:BR}).
For $\tau \to \mu  \bar{\nu}  \nu \gamma$ decays, the branching ratio measurement and prediction agree within 1.1 standard deviations (1.1$\sigma$). On the contrary, the experimental and theoretical values for $\tau \to e \bar{\nu}  \nu \gamma$ decays differ by $2.02 \, (57) \times 10^{-3}$, i.e.\  by 3.5$\sigma$. This puzzling discrepancy deserves further researches. From table~\ref{tab:BR} we also note that the inclusive branching ratio of $\tau \to e  \bar{\nu}  \nu \gamma$ is less sensitive to the NLO corrections than the exclusive one.

The branching ratio of the radiative decay $\mu \to e \bar{\nu} \nu \gamma$ was measured long ago for a minimum photon energy $\omega_0=10$~MeV in the $\mu$ rest frame~\cite{Crittenden:1959hm},
\begin{equation}
   \mathcal{B}_{\scriptscriptstyle \rm EXP} \left(\mu \to e  \bar{\nu}  \nu \gamma  \right) 
   = 1.4  \, (4) \times 10^{-2}.
   \label{eqn:MUE1961}
\end{equation}
This measurement agrees with our theoretical prediction, and new precise results are expected to be published in the near future by the \textsc{Meg} and \textsc{Pibeta} collaborations~\cite{Adam:2013gfn,Pocanic:2014mya}.
%
\begin{table}[tbp]
\def\arraystretch{1.2}
   \centering
   \begin{tabular}{|l|l|l|l|}
     \hline &
	$\tau \to e \bar{\nu} \nu \gamma$ 	&
	$\tau \to \mu \bar{\nu} \nu \gamma$	&
	$\mu \to e \nu \bar{\nu}\gamma$ 	\\
      	\hline
	$\mathcal{B}_{\scriptscriptstyle \rm LO}$ 				&	
	$ 1.834 \times 10^{-2}$ &
      	$ 3.663 \times 10^{-3}$ &
 	$ 1.308 \times 10^{-2}$ \\
	$\mathcal{B}_{\scriptscriptstyle \rm NLO}^{\rm Inc}$ 		&
      	$-1.06  \, (1)_n (10)_N \times 10^{-3}$ &     
      	$-5.8  \, (1)_n (2)_N \times 10^{-5}$ &     
      	$-1.91  \, (5)_n (6)_N \times 10^{-4}$ \\     
	$\mathcal{B}_{\scriptscriptstyle \rm NLO}^{\rm Exc}$	&
	$-1.89  \, (1)_{n} (19)_N \times 10^{-3}$ &     
      	$-9.1  \, (1)_{n} (3)_N \times 10^{-5}$ &     
      	$-2.25  \, (5)_{n}  (7)_N \times 10^{-4}$ \\     
	$\mathcal{B}^{\rm Inc}$							&
      	$ 1.728  \, (10)_{\rm th} (3)_{\rm \tau} \times 10^{-2}$ &      
      	$ 3.605  \, (2)_{\rm th} (6)_{\rm \tau} \times 10^{-3}$ &      
      	$ 1.289  \, (1)_{\rm th} \times 10^{-2}$ \\      
	$\mathcal{B}^{\rm Exc}$							&
      	$ 1.645  \, (19)_{\rm th} (3)_{\rm \tau} \times 10^{-2}$ &      
      	$ 3.572  \, (3)_{\rm th} (6)_{\rm \tau} \times 10^{-3}$ &      
      	$ 1.286  \, (1)_{\rm th} \times 10^{-2}$ \\     
 	$\mathcal{B}_{\scriptscriptstyle \rm EXP}$				&
      	$ 1.847  \, (15)_{\rm st} (52)_{\rm sy} \times 10^{-2}$ &      
      	$ 3.69  \, (3)_{\rm st} (10)_{\rm sy} \times 10^{-3}$ &      
      	$ 1.4  \, (4) \times 10^{-2}$ \\     
\hline
    \end{tabular}
\caption{Branching ratios of radiative $\mu$ and $\tau$ leptonic decays with minimum photon energy $\omega_0 = 10$~MeV. Inclusive ($\mathcal{B}^{\rm Inc}$) and exclusive ($\mathcal{B}^{\rm Exc}$) predictions are separated into LO contributions ($\mathcal{B}_{\scriptscriptstyle \rm LO}$) and NLO corrections ($\mathcal{B}_{\scriptscriptstyle \rm NLO}^{\rm Inc/Exc}$). Uncertainties were estimated for uncomputed NNLO corrections ($N$), numerical errors ($n$), and the experimental errors of the lifetimes ($\tau$). The first two types of errors were combined to provide the final theoretical uncertainty (th). The last line reports the experimental measurements of refs.~\cite{Lees:2015gea,Crittenden:1959hm}.}
\label{tab:BR}
\end{table}

\section{Conclusions}\label{sec:conclusions}

In this work we studied the SM prediction of the differential rates and branching ratios of the radiative decays $\tau \to l \bar{\nu} \nu \gamma$ $(l=e,\mu)$ and $\mu \to e \bar{\nu} \nu \gamma$. 
The NLO corrections were computed using the effective four-fermion Fermi Lagrangian plus QED, taking into account the full depencence on the mass ratio $r=m/M$. The resulting differential rates for the radiative decays of a polarized $\mu$ or $\tau$ were presented in section~\ref{sec:nlo}, eq.~\eqref{eqn:radiativedecayrateNLO}. There, the functions $G$, $J$, $K$, and $L$ contain the LO contributions, inclusive of tiny effects induced by the $W$-boson propagator, the virtual corrections at NLO, and the "soft" double bremsstrahlung decay rate due to events in which one of the two photons is too soft to be detected. Explicit analytic expressions for $G$, $J$, $K$, and $L$ are provided in attachment to this paper. The differential rate of the leptonic decay of an unpolarized $\mu$ or $\tau$ with "hard" double-photon emission, necessary to evaluate the branching ratios, was also calculated. Agreement was found with an earlier calculation of the isotropic function $G$, which was however performed in the $r \to 0$ limit. On the contrary, our anisotropic $J$ and $K$ functions differ from an earlier calculation, even in the massless $r \to 0$ limit, and the anisotropic function $L$ has been previously overlooked.

The branching ratios for a minimum photon energy $\omega_0 = \ymin (M/2)$ were presented in section~\ref{sec:br} integrating the differential rates of section~\ref{sec:nlo} over the allowed kinematic ranges. Particular attention was paid to terms proportional to $r^2$ in the integrand $G$, as they lead to terms of relative $\mathcal{O} (1)$ in the integrated result -- they cannot be neglected. Branching ratios were distinguished in "exclusive", $\mathcal{B}^{\rm Exc} \left( y_0 \right)$, measured counting the decays with the emission of one, and only one, photon of energy larger than the detection threshold $y_0$, and "inclusive" ones, $\mathcal{B}^{\rm Inc} \left( y_0 \right)$, defined measuring radiative decays with the emission of at least one photon with normalized energy higher than $\ymin$. Precise numerical results were presented for both of them, in table~\ref{tab:BR}, for $\omega_0 = 10$~MeV. Uncertainties were estimated for uncomputed NNLO corrections, numerical errors, and the experimental errors of the lifetimes.

Our predictions agree with the old measurement of the branching ratio $\mathcal{B} (\mu \to e \bar{\nu} \nu \gamma)$, and precise new results for this decay are expected to be published in the near future by the \textsc{Meg} and \textsc{Pibeta} collaborations.
Also the recent precise measurement by \textsc{Babar} of the branching ratio $\mathcal{B} (\tau \to \mu  \bar{\nu} \nu \gamma)$, for $\omega_0 = 10$~MeV, agrees with our prediction within 1.1 standard deviations (1.1$\sigma$). On the contrary, \textsc{Babar}'s recent measurement of the branching ratio $\mathcal{B} (\tau \to e \bar{\nu}  \nu \gamma)$, for the same threshold $\omega_0$, differs from our prediction by 3.5$\sigma$.  This puzzling discrepancy deserves further researches.

\acknowledgments
We would like to thank A.~Arbuzov, M.~Giorgi, A.~Lusiani, P.~Nason, B.~Oberhof, F.~Piccinini, A.~Sirlin, R.~Torre, Z.~Was, and A.~Wulzer for very useful discussions and correspondence. 
We would also like to thank D.~Epifanov and S.~Eidelman for our fruitful and ongoing collaboration.
The work of M.F.\ and L.M.\ is supported by the Swiss National Science Foundation.
M.P.\ also thanks the Department of Physics and Astronomy of the University of Padova for its support. His work was supported in part by the Italian Ministero dell'Universit\`a e della Ricerca Scientifica under the program PRIN 2010-11, and by the European Program INVISIBLES (contract PITN-GA-2011-289442).



\end{document}